\documentclass[usenatbib]{mnras}

\usepackage{newtxtext,newtxmath}
\usepackage[T1]{fontenc}

\DeclareRobustCommand{\VAN}[3]{#2}
\let\VANthebibliography\thebibliography
\def\thebibliography{\DeclareRobustCommand{\VAN}[3]{##3}\VANthebibliography}

\usepackage{graphicx}	% Including figure files
\usepackage{amsmath}	% Advanced maths commands
\usepackage{amssymb}	% Extra maths symbols
\usepackage{hyperref}
\usepackage{cleveref}
\title[M10: stellar membership, age and distance]{The globular cluster M10: Reassessment of stellar membership, distance and age using its variable and HB stars $\thanks{Based on observations made at the observatories Indian Astrophysical Observatory, Hanle (India), San Pedro M\'artir Observatory, (Mexico) and Bosque Alegre (Argentina).}$}

\author[Arellano Ferro et al.]{
A. Arellano Ferro,$^{1}$\thanks{E-mail: armando@astro.unam.mx} M. A. Yepez,$^{1}$ S. Muneer,$^{2}$
I.H. Bustos Fierro,$^{3}$ K.P. Schr\"oder,$^{4}$ \and
Sunetra Giridhar,$^{2}$
J.H. Calder\'on,$^{3,5}$
\\
$^{1}$Instituto de Astronom\'ia, Universidad Nacional Aut\'onoma de M\'exico, Ciudad de M\'exico, CP 04510, M\'exico.\\
$^{2}$Indian Institute of Astrophysics, Bangalore, India\\
$^{3}$Universidad Nacional de C\'ordoba. Observatorio
Astron\'omico. C\'ordoba,  Argentina.\\
$^{4}$Departamento de Astronom\'ia, Universidad de Guanajuato, M\'exico.\\
$^{5}$Consejo Nacional de Investigaciones Cient\'ificas y
T\'ecnicas (CONICET), Buenos Aires, Argentina.\\
}

\date{Accepted: ; Received: 2020  in original form 2020 July}

\pubyear{2020}

\begin{document}
\label{firstpage}
\pagerange{\pageref{firstpage}--\pageref{lastpage}}
\maketitle

% Abstract of the paper
\begin{abstract}
Time-series \emph{VI} CCD photometry of the globular cluster M10 (NGC 6254) is employed to perform a detailed identification, inspection of their light curves, their classification and their cluster membership, of all the known variables reported up to 2018. The membership analysis is based on the $Gaia$-DR2 positions and proper motions. The metallicity of the cluster is estimated based on the sole RRc star known in the cluster. The Fourier decomposition of its light curve leads to [Fe/H]$_{\rm ZW}$=$-1.59
\pm 0.23$ dex. The mean cluster distance, estimated by several independent methods is $5.0 \pm 0.3$ kpc.
A multi-approach search in a region of about 10$\times$10 arcmin$^2$ around the
cluster revealed three new variables, one SX Phe (V35) and two sinusoidal variables on the red giant branch of unclear classification (V36,V37).
Modelling the HB stars is very sensitive to the stellar hydrogen shell
mass, which surrounds the 0.50 $M_{\odot}$ helium core. To match the full
stretch of the HB population, a range of total mass of 0.56 to 0.62
$M_{\odot}$ is required. These models support a distance of 5.35 kpc and an age of about 13 Gyrs, and hints to some individual variation of the mass loss on the uppper
RGB, perhaps caused by the presence of closed magnetic field in red
giants.
\end{abstract}

% Select between one and six entries from the list of approved keywords.
% Don't make up new ones.

%%%%%%%%%%%%%%%%%%%%%%%%%%%%%%%%%%%%%%%%%%%%%%%%%%

%%%%%%%%%%%%%%%%% BODY OF PAPER %%%%%%%%%%%%%%%%%%
%\begin{document}
% Typeset article header
\maketitle

\section{Introduction}
\label{sec:intro}

The globular cluster M10 (NGC 6254; C1654-040 in the IAU nomenclature) ($\alpha = 16^{\mbox{\scriptsize h}}
57^{\mbox{\scriptsize m}} 09.05^{\mbox{\scriptsize s}}$, $\delta =-04^{\mbox{\scriptsize o}} 06{\mbox{\scriptsize '}} 01.1{\mbox{\scriptsize "}}$,
J2000; $l = 15.14^{\mbox{\scriptsize o}}$, $b = +23.08^{\mbox{\scriptsize o}}$), was neglected for a long time by the researchers interested on the variable star populations in globular clusters (GC). The first known variables, V1 and V2 were discovered by \citet{Sawyer1938}, and V3 and V4 by \citet{Arp1955a,Arp1955b}. It was going to take more than 60 years, already in the CCD era, that numerous variables, V5-V16 were to be discovered by \citet{Salinas2016} (hereinafter SA16), most of them faint stars in the central region of the cluster. More recently \citet{Rozyczka2018} (hereinafter RO18), explored the outskirts of the cluster to find  another substantial group of variables considered cluster members, V17-V34, as well as some non-members in the field of the cluster, labeled N1-N6.

In the present paper we report the results of the analysis of a new time-series of \emph{VI} CCD images, aimed to  confirm the variability and classifications  of the variable star
population, search for new variables, and to discuss  their cluster membership.

The confirmed and suspected variables presently known in the field of M10,  after the results of the present paper are considered, is formed by 41 stars; 16 SX Phe, 2 W Vir, 1 BL Her, 2 EW, 1 EA, 2 RRc, 5 SR type stars, and 11 unclassified stars with sinusoidal clear or suspected variations, and 1 star classified as $\delta$ Scuti (N1 of RO18).  
Stars identified as V4  and V16 show no signs of variability (RO18). V4 is out of the field of our images.
The field star status of several of the above variables is clear or suspected as shall be discussed in the present work.

No proper identification charts have been published, which makes the identification of some variables difficult or dubious. We have found some inconsistencies between the identifications of SA16 and RO18 and will make an attempt to clarify them, since, keeping our knowledge of the variable star population in the Galactic globular cluster system tidy, complete and properly classified, would help achieving the fundamental goal of observational astronomy, i.e. transforming observational quantities into physical parameters. It is in this scope that the present paper is framed.

In this paper we describe our observations and data reductions as well as the
transformation to the Johnson-Kron-Cousins photometric system ($\S$ 2), we perform 
a search of new variables in our collection of light curves of stars in the FoV of the cluster, and report three new ones ($\S$ 3). Using the proper motions in the $Gaia$-DR2 and the method of \citet{Bustos2019}, we separate cluster members from field stars and discuss their effects on the Colour-Magnitude diagram (CMD) and the membership status of the variable stars ($\S$ 4). A detailed 
identification of all known variables is carried out and the  preparation of light curves is discussed ($\S$ 5). The Fourier decomposition of the sole RRc star in the cluster is discussed ($\S$ 6). We present a CMD, cleaned from  field stars, where we position the variable stars to confirm their classification. Isochrones suggest the cluster being older than 13 Gyrs ($\S$ 7). The period-luminosity relationship for SX Phe stars is presented and used to identify the pulsation mode
of the main frequency and the mean cluster distance ($\S$ 8). The major signals in the frequency spectra of the SX Phe stars are listed and the pulsation modes are identified whenever possible ($\S$ 9). As the blue tail of the Horizontal Branch (HB) is a very sensitive
indicator of the remaining shell mass of stars in central Helium burning,
and in M10 it is quite populated, we used the evolutionary code described by \citet[][]{Pols1997, Pols1998}  and \citet{KPS1997}, to model the HB and discuss the implications of the inferred mass loss on the RGB and the distance and age of the cluster ($\S$ 10). Finally we  
summarize our results ($\S$ \ref{sec:Summ}). In Appendix A, we discuss the properties and
classification of a number of peculiar variables that may require further analysis.

\section{Observations and reductions}
\label{sec:ObserRed}

\begin{table}
\footnotesize
\caption{The distribution of observations of M10.$^*$}
\centering
\begin{tabular}{lcccccc}
\hline
Date  &  $N_{V}$ & $t_{V}$ & $N_{I}$ &$t_{I}$&Avg &site\\
      &          & sec    &         &sec&seeing (")&\\
\hline
 2018-06-18 & 54 & 60 & 68 & 40   & 1.8&SPM\\
 2018-07-12 & 20 & 60 & 21 & 40  & 1.5&SPM\\
 2018-07-13 &  2 & 60 &  -- & --  & 1.8&SPM\\
 2018-07-14 & 52 & 60 & 57 & 40 & 1.8&SPM\\
 2018-07-24 & 2 & 60 & 5 & 40 & 1.6&SPM\\
 2018-08-11 &  4 & 60 &  3 & 30 & 2.6&BA\\
 2018-09-02 & 20 & 60 & 20 & 30 & 2.2&BA\\
 2018-09-14 & 34 & 60  & 36 & 30 & 2.2&BA\\
 2019-05-26 & 6 & 60 & 25 & 40 & 2.3&SPM\\
 2019-05-27 & 2 & 60 & 3 & 40 & 3.2&SPM\\
 2019-05-28 & 23 & 60 & 55 & 40  & 2.5&SPM\\
 2019-05-29 &  12 & 60 & 60 & 40 & 2.2&SPM\\
 2019-05-30 & 4 & 60 & 37 & 40 & 2.5&SPM\\
 2019-06-25 & 41 & 60 & 52 & 40 & 2.1&SPM\\
 2019-06-26 & 30 & 60 &  59 & 40  & 1.7&SPM\\
 2019-06-27 & 52 & 60 & 63 & 40 & 1.9&SPM\\
 2019-06-28 & 21 & 60 & 28 & 40 & 2.0&SPM\\
 2019-06-30 & 4 & 60 & 26 & 40 & 1.7&SPM\\
 2019-07-01 & 40 & 60 & 60 & 40 & 1.7&SPM\\ 
 2019-09-01 & 18 & 80 & 18 & 40 & 3.3&BA\\
 2019-09-02 & 34 & 80 & 37 & 40 & 2.9&BA\\
 2020-04-24 & 80 & 10 & 80 & 30 & 1.3&IAO\\
\hline
Total:   & 555&    &  813  &    & &\\
\hline
\end{tabular}
\raggedright
\quad $*$Columns $N_{V}$ and $N_{I}$ give the number of images taken with the $V$ and $I$
filters respectively. Columns $\MakeLowercase{t}_{V}$ and $\MakeLowercase{t}_{I}$
provide the exposure time,
or range of exposure times. In the last two columns the average seeing and the observatory are listed.
\label{tab:obs}
\end{table}

\subsection{Observations}

The Johnson-Kron-Cousins $V$ and $I$ observations 
used in the present work were obtained from three sites; between  June 2018 and July 2018 and between May 2019 and July 2019 with 
the 0.84m-telescope at the of the San Pedro M\'artir observatory, in Baja California, M\'exico. The detector in 2018 was a Spectral Instruments CCD of 1024$\times$1024 pixels with a scale of 0.444 arcsec/pix, translating to a field of view (FoV) of approximately 7.57$\times$7.57~arcmin$^2$.
In 2019 the detector was 1024x1032 pixels with a scale of 0.493 arcsec/pixel and a field of 
$8.41\times8.48$ arcmin$^2$.

The second site was the Bosque Alegre Astrophysical Station
of the C\'{o}rdoba Observatory, Universidad Nacional de C\'{o}rdoba,
Argentina, with the
1.54-m telescope.
Observations were acquired 
between August and September 2018 and in September 2019.
The detector in 2018 was a CCD KAF-16803 of 4096x4096 pixels, and in 2019 a CCD KAF-6303E of 3072x2048 pixels. The FoV on the first CCD was 16.9 x 16.9 arcmin$^2$, and on the second one 12.6 x 8.4 arcmin$^2$. In both seasons the detectors were binned 2 x 2 for a scale of 0.496 arcsec/pixel, and the frames were trimmed to a maximum size of approximately 10 arcmin$^2$ in order to avoid images severely affected by coma.

%\begin{figure*}
%transformation\begin{center}
%transformation\includegraphics[width=16cm]{MOSAI_TRANSF.pdf}
%\caption{The transformation relationships between the instrumental and standard
%photometric systems for the runs SPM2018 (left panel) and IAO2020 (right panel), using a set of 141 and 195 standards respectively from the list of \citet{Stetson2000} in the FoV of our images of M10. A small but significant colour term is evident.}
%    \label{trans}
%\end{center}
%\end{figure*}

The third site was the Indian Astronomical Observatory (IAO) in Hanle, India, where observations were acquired
with the 2 m telescope on April 24, 2020. The detector used was a E2V CCD44-82-0-E93 of $2048\times4096$
pixels with a scale of 0.296 arcsec/pix. A sector of  $2048\times2048$ pixels was used,
translating to FoV of approximately $10.1\times10.1$ arcmin$^2$.

Table \ref{tab:obs}
gives an overall summary of our observations and the seeing conditions. For clarity purposes, in what follows we shall refer to these five runs as SPM2018, SPM2019, BA2018, BA2019 and IAO2020.

\subsection{Difference Image Analysis}
\label{DIA}

Image data were calibrated using bias and flat-field
correction procedures. We used the Difference Image Analysis (DIA)
to extract high-precision time-series photometry in the FoV of M10. We used
the 
{\tt DanDIA}\footnote{{\tt DanDIA} is built from the DanIDL library of IDL routines
available at \texttt{http://www.danidl.co.uk}}
pipeline for the data reduction process \citep{Bramich2013}, which includes an 
algorithm that models the convolution kernel matching the PSF
of a pair of images of the same field as a discrete pixel array \citep{Bramich2008}. 
A detailed description of the procedure is available in the paper by \citet{Bramich2011}, to which the interested reader is referred for 
the relevant details.

We also used the methodology developed by \citep{Bramich2012} to solve for the 
magnitude offset that may be introduced into the photometry 
by the error in the fitted value of the photometric scale factor
corresponding to each image.  
In the present case, the magnitude offset due to this error was rather negligible, of the order
of $\approx 0.5{-}1.0$~mmag for stars brighter than $\sim$ 18.0~mag.

\subsection{Transformation to the \textit{VI} standard system}

We made use of local standar stars,
gathered from the  standar star  collection of \citep{Stetson2000}\footnote{%
\texttt{http://www3.cadc-ccda.hia-iha.nrc-cnrc.gc.ca/\\
community/STETSON/standards}}, which have been set into the Johnson-Kron-Cousins standard system using the equatorial standards from  \citet{Landolt1992}.

As can be appreciated in Table \ref{tab:obs}, the seasons SPM2018 and IAO2020 were the ones that produced the best quality images, hence we use those seasons to set our observations into the standards system. The transformation equations  show a small scatter ($\sim$ 0.005-0.008 mag) and carry a small colour dependence. For the other seasons, due mainly to poor seeing conditions and crowding for a good fraction of the local standards, small zero point differences were present, and were simply applied to bring the magnitude system to match the IAO2020 level.

In Table \ref{tab:vi_phot} we include a small portion of the time-series \emph{VI} photometry obtained in this work. The full table shall be available in electronic form in the Centre de Donnés astronomiques de Strasbourg database (CDS).

\begin{table}
\scriptsize
\begin{center}
\caption{Time-series \textit{VI} photometry for the variables stars observed in this work$^*$}
\label{tab:vi_phot}
\centering
\begin{tabular}{cccccc}
\hline
Variable &Filter & HJD & $M_{\mbox{\scriptsize std}}$ &
$m_{\mbox{\scriptsize ins}}$
& $\sigma_{m}$ \\
Star ID  &    & (d) & (mag)     & (mag)   & (mag) \\
\hline
 V1 & $V$& 2458342.52058& 11.849 & 14.402 & 0.001 \\   
 V1 & $V$& 2458342.52211& 11.849 & 14.402 & 0.001 \\
\vdots   &  \vdots  & \vdots & \vdots & \vdots & \vdots  \\
 V1 & $I$ & 2458342.51828 & 10.260 & 13.681 & 0.001\\  
 V1 & $I$ & 2458342.51870 & 10.256 & 13.677 & 0.001  \\ 
\vdots   &  \vdots  & \vdots & \vdots & \vdots & \vdots  \\
 V2 & $V$ & 2458727.54934 & 12.591& 15.755 & 0.002 \\   
 V2 & $V$ & 2458727.55030 & 12.591& 15.755&  0.001 \\
\vdots   &  \vdots  & \vdots & \vdots & \vdots & \vdots  \\
 V2 & $I$ & 2458727.54759 & 11.381&  14.984 & 0.002 \\    
 V2 & $I$ & 2458727.54809 & 11.387&  14.991 & 0.001 \\   
\vdots   &  \vdots  & \vdots & \vdots & \vdots & \vdots  \\
\hline
\end{tabular}
\end{center}
* The standard and
instrumental magnitudes are listed in columns 4 and~5,
respectively, corresponding to the variable stars in column~1. Filter and epoch of
mid-exposure are listed in columns 2 and 3, respectively. The uncertainty on
$\mathrm{m}_\mathrm{ins}$, which also corresponds to the
uncertainty on $\mathrm{M}_\mathrm{std}$, is listed in column~6. A full version of this table is available at the CDS database.

\end{table}

\section{Search for new variables}

\begin{figure}
\begin{center}
\includegraphics[width=8cm]{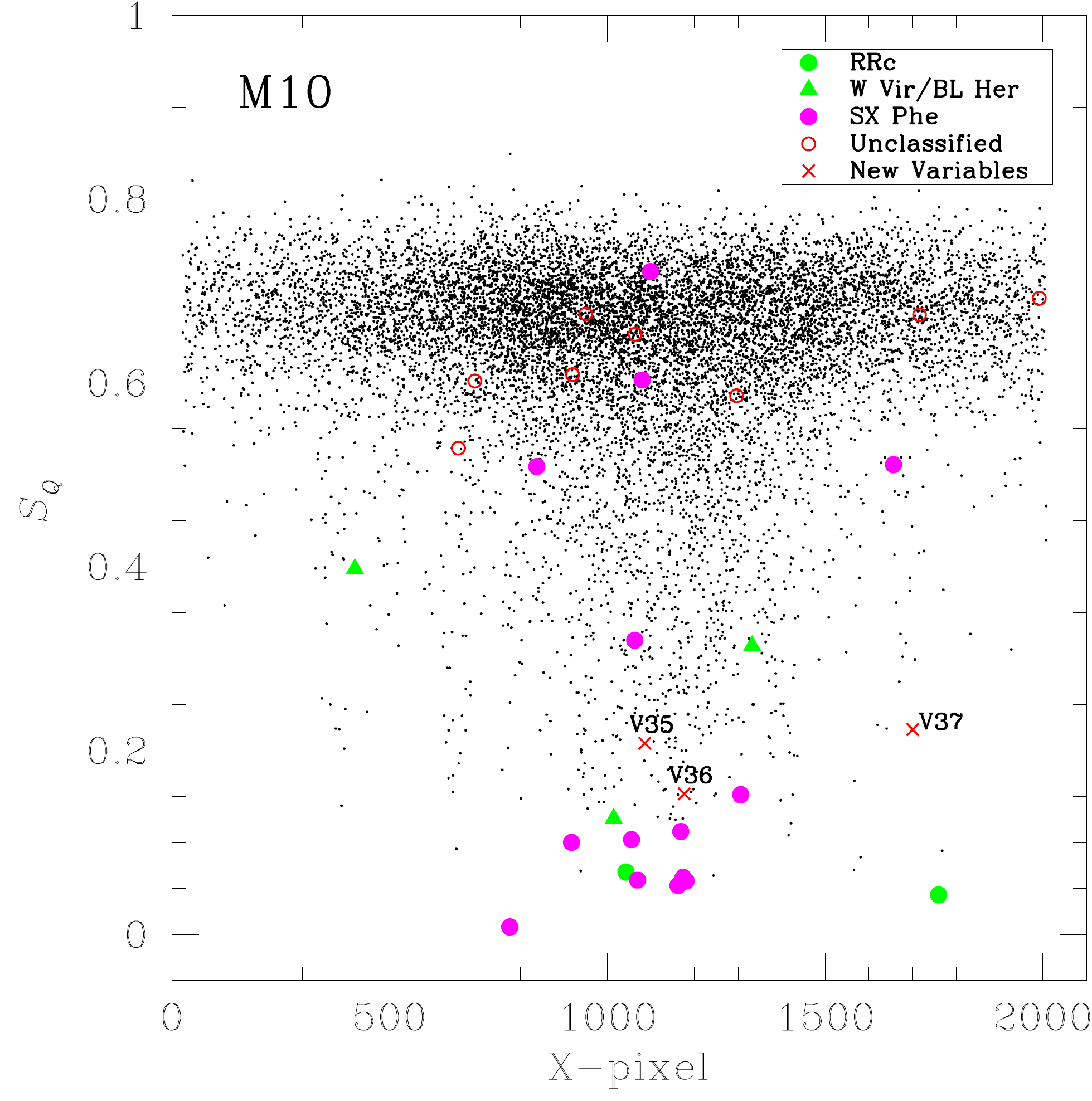}
\caption{The distribution of the string length statistic $S_Q$ versus in all measured stars in the field of M10 in the IAO2020 season. Known variables are coloured following the code in the legend.}
    \label{SQ}
\end{center}
\end{figure}

\begin{figure*}
\begin{center}
\includegraphics[width=18.cm]{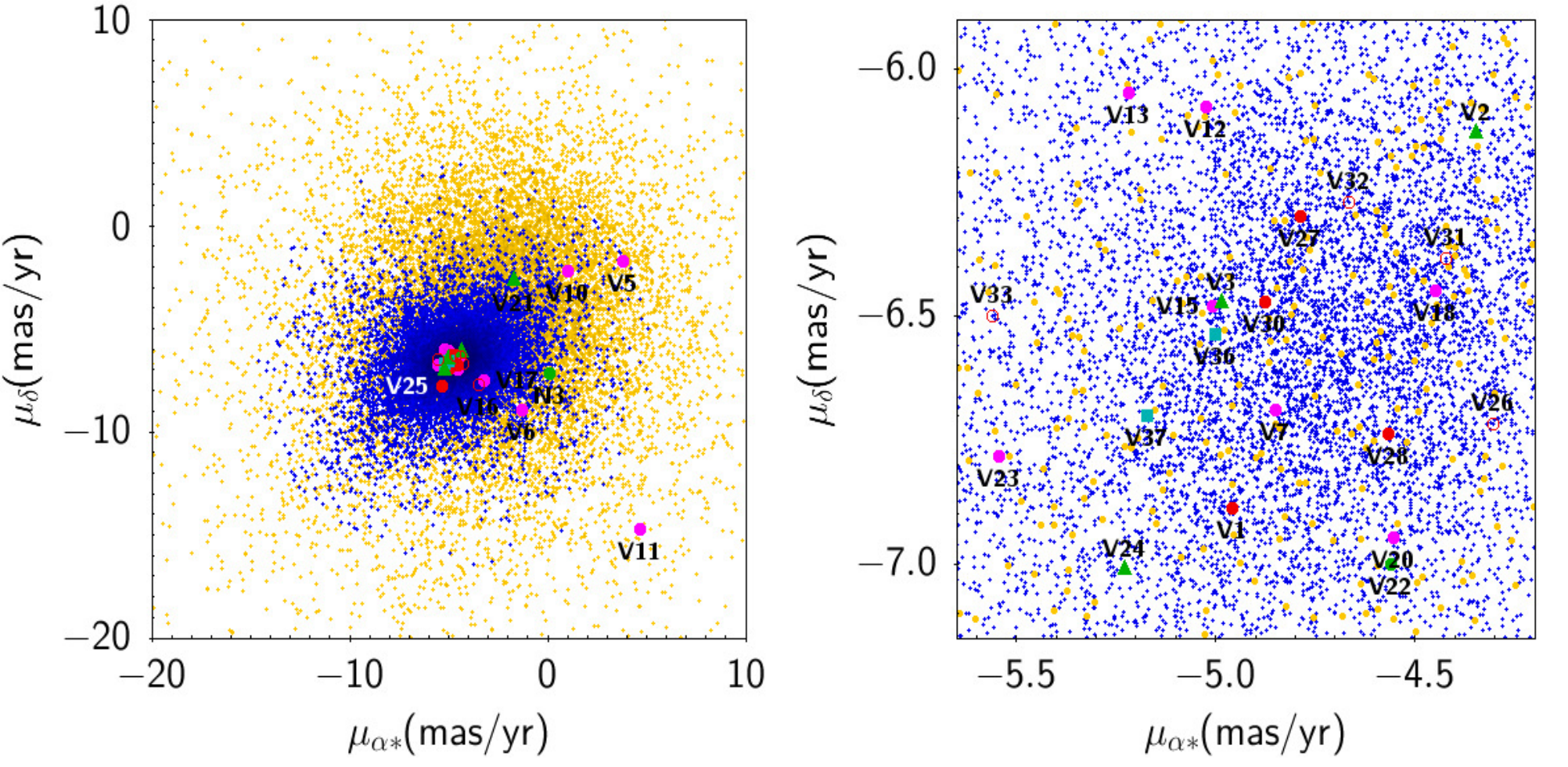}
\caption{VPD of the cluster M10. Blue and yellow points are stars that were found cluster members or non-members respectively ($\S$ \ref{membership}). Variable star are identified following the colour code of Fig. \ref{SQ}. Right panel shows the central region.}
    \label{VPD}
\end{center}
\end{figure*}

A detailed search was conducted on the IAO2020 data. Two approaches were followed: first, the string length method (\citealt{Burke1970}, \citealt{Dworetsky1983}) (or see \citet{Deras2020} for a detailed description of the method), that assigns a statistical indicator $S_Q$, of the dispersion of individual observations in a given light curve phased with a trial period. The minimum value of $S_Q$ is obtained when the dispersion is least and hence the period likely the correct one.
The method is prone to spurious results if the scatter of the light curves is large. Since our best quality data are those from the IAO2020 season, we limited the analisis to those data, which, however, span only 7 hours. Hence, we are aiming to identify stars with periods shorted that 0.3 days.
The distribution of $S_Q$ plotted versus the X-coordinate of each star is displayed in Fig. \ref{SQ}. Large amplitude and/or low dispersion light curves tend to fall in the lower part of the diagram, hence, exploring the light curves individually below a given threshold, offers a good chance to find variables previously undetected.  Therefore, we visually explored all the light curves below $S_Q = 0.5$, phased with the best candidate period, searching for convincing variability. We were able to recover all the previously known variables but found no new ones. 

The second approach was performed throughout the blinking of the 80 $V$ residual images from the IAO2020 season. This method confirmed all the already known variables and pointed to  three new variables now labeled V35-V37. Their periods, light curves and classifications are included in subsequent tables and figures and they are discussed later in the paper.

\section{On the cluster membership of stars in the FoV}
\label{membership}

Using the high-quality astrometric data available in Gaia-DR2 \citep{Gaia2018}  and the method of \citet{Bustos2019}, we were able to separate 21717 likely cluster members from a total of 36692 stars with Gaia-DR2 proper motions, in the field of radius 25 arcmin centered in the cluster. The corresponding Vector Point Diagram (VPD) is shown in Fig, \ref{VPD}, where it can be seen that the distribution of the proper motions of field and cluster stars (black and blue symbols in the figure) are not well separated, which implies that some  contamination by field stars of our member stars list is likely to be present. We possess light curves for 9249 stars, which shall be used to build a cleaner version of the CMD of the cluster. The membership status of all variable stars, based on their position on the VPD and CMD will be addressed later in Appendix A.

\section{Variable Stars light curves in M10}

To further study the variable stars in M10, an accurate identification in the field of the cluster and a confirmation of their variability in our data are in order. The equatorial coordinates for each variable were taken from SA16 and RO18. A few inconsistencies were detected and we have made every effort to confirm 
the variability of the star. 
We based our identifications
on the images and data from the IAO2020 season, which has the best seeing conditions, and have been able to recover the variability of all known variables in the FoV of our images.

All the variable stars in our FoV, are listed in Table \ref{tab:variables} with some of their basic data, i.e. mean intensity weighted magnitudes, amplitudes, periods and coordinates and membership status.

 Their light curves, as evidenced in our observations, are displayed in Fig. \ref{VariablesA}. All the light curves are phased with the period rendered by our own observations, given in column 7, otherwise the period of RO18 in column 8 was used. We were unable to detect variations in star V16 with a declared amplitude $\sim  0.03$ mag and  whose variability was detected only in 2015 (RO18).

\begin{figure*}
\begin{center}
\includegraphics[scale=1.8]{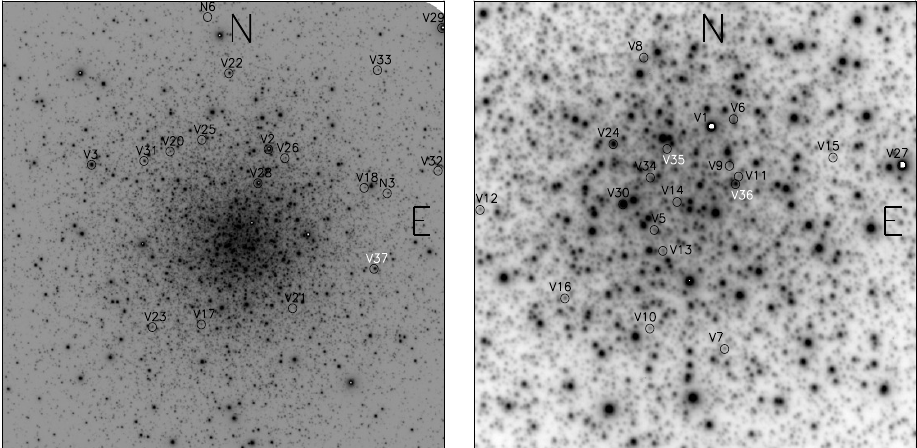}
\caption{Finding chart constructed from the $V$ reference images for the IAO2020 season. The left panel
displays the complete FoV of our images and it is about $10.1\times10.1$~arcmin$^{2}$.
The panel on the right displays the central region of the cluster and it is about
$2.9\times2.9$~arcmin$^{2}$. All stars listed in Table
\ref{tab:variables} are identified in at least one of the panels. Variables labeled in white, are newly identified in this work.}
\label{chart}
\end{center}
\end{figure*}

The finding chart of all
variables in
Table~\ref{tab:variables} is shown in Fig.~\ref{chart} for the cluster peripheral and core regions.
We employed for that purpose, the coordinates reported in table 1 of RO18, on our reference image for the IAO2020 season, duly  calibrated astrometricaly. Unfortunately, that table contains some errors in the coordinates of a few variables, which have been kindly clarified and corrected by Dr. M. Rozyczka. In Table~\ref {tab:variables} we include the correct coordinates for the variables and recommend
their usage in future studies of these stars. In Fig. \ref{chart} we also include the new discoveries in this work V35, V36 and V37. A discussion of peculiarities of specific stars and membership status shall be deferred to the Appendix A at the end of this work.

\begin{table*}
%\scriptsize
\begin{center}
\caption{Data of Variable stars in M10 in the F\lowercase{o}V of our images.}
\label{tab:variables}
\begin{tabular}{llcccclccccc}
\hline
ID & Variable & $<V>^1$ & $<I>^1$ & $A_V$ & $A_I$ & $P^2$ (days) & $P$ (days) & HJD$_{\rm max}$ &  RA  &  Dec. & mem$^3$\\
  &  Type  & (mag) & (mag) & (mag) & (mag) & this work & RO18 & (2450000+)  & (J2000.0) & (J2000.0)&\\
\hline
V1  & SR           & $\it 11.809$ & $\it 10.226$ & --& -- & -- & 70.878903  & -- & 16:57:10.11&--4:05:36.10 &Y \\
V2  & W Vir & $\it 12.127$ & $\it 10.934$ &  -- & -- &  -- & 19.470995 & -- & 16:57:11.74&--4:03:59.69&Y\\
V3  & W Vir           & 12.761 &  11.721   & 0.34 &   0.36  &  7.835134 & 7.872181 &  8342.5737  & 16:56:55.96 &--4:04:16.43  &Y\\
V5  & SX  Phe & 17.079 & 16.600 & 0.53 & 0.37 &  0.058550& 0.058543 & 8662.8713 & 16:57:08.59&--4:06:16.31 &Y \\
V6  & SX Phe & 16.717 & 16.104 & 0.09 & 0.03 &  0.063731 & 0.059909 & 8313.7535 & 16:57:10.70&--4:05:33.36&Y \\
V7  & SX Phe  & 17.592 & 16.900 & 0.10 & 0.09 &  0.048106 & 0.048112 & 8311.8444 & 16:57:10.37 &--4:07:03.29 &Y \\
V8  & SX Phe& 17.012 &16.339& 0.10  &0.08& 0.051007& 0.051009& 8311.8168 & 16:57:08.38&--4:05:08.74&?  \\
V9  & SX Phe& 17.303 & 16.166 & 0.60 & 0.29 &  0.051312 & 0.051301 & 8964.3137& 16:57:10.57&--4:05:51.79 &?\\
V10 & SX Phe& 17.555 & 17.048 & 0.10 & -- &  0.02259 & 0.022319 & 8313.8231 & 16:57:08.43&--4:06:54.79 &Y  \\
V11 & SX Phe& 17.515 & 16.740 & 0.265 & 0.131 & 0.047958  & 0.047957  & 8964.4079 & 16:57:10.82&--4:05:55.90 &N\\
V12 & SX Phe & 17.305  & 16.649 & 0.04 &-- &  -- & 0.022823 & 8964.3572 & 16:57:04.05&--4:06:07.31 &Y\\
V13 & SX Phe& 16.896 & 16.369 & 0.04 & 0.03 &   0.06495 & 0.036944 & 8964.3142 & 16:57:08.80&--4:06:24.48 &Y \\
V14 & SX Phe& 17.641 & 16.912 & 0.10 & 0.07 &  0.041245  & 0.038198 & 8964.4016&
16:57:09.19&--4:06:05.36&?\\
V15 & SX Phe& 17.496 & 16.838 & 0.07 & 0.06 &  0.034835 & 0.034835 & 8313.8457 & 16:57:13.28&--4:05:48.98 &Y \\
V16 & No var & 16.904 & 15.696 & -- & 0.02 &  0.357809 & 0.357809 & -- & 16:57:06.23&--4:06:42.52 &Y \\
V17 & SX Phe& 17.284 & 16.600 & 0.10 & 0.10 &  0.036946 & 0.036944 & 8964.4153 & 16:57:05.52&--4:07:47.32 &Y\\
V18 & SX Phe& 17.534  &16.938  & 0.07 & 0.07 & 0.041090 & 0.042435 & 8964.3371 & 16:57:20.23&--4:04:52.18&Y\\
V19 & SX Phe& -- & -- & -- & -- &  -- & 0.043795 & -- & 16:57:38.66&--4:08:57.41  &Y\\
V20 & SX Phe& 16.987 & 16.463 & 0.59 & 0.40 & 0.050603  & 0.050603 & 8287.7290 &16:57:02.97&--4:04:00.59&Y\\
V21 & EW  & $\it 19.682$ & $\it 18.434$ &0.32 & 0.29 &  -- & 0.244976 & -- & 16:57:13.69&--4:07:28.13&Y \\
V22 & RRc   & 14.637 & 13.974 & 0.39 & 0.26 &  0.404485 & 0.404604 & 8313.7235 & 16:57:08.32&--4:02:19.79&Y \\
V23 & sin & 17.657 & 16.869 & 0.09 & -- & -- & 1.446583 & 8287.8697 & 16:57:01.15&--4:07:49.73 &Y\\
V24 & BL Her& 14.023 & 12.944 & 0.353 & 0.300 &  -- & 2.307458 & 8287.7232 &16:57:07.55&--4:05:42.36&Y \\
V25 &  sin  & 17.322 & 17.185 & 0.07 & 0.11 & -- & 4.457001 & -- & 16:57:05.83&--4:03:46.04 &? \\
V26 & sin & 16.573 & 15.287 & 0.18 & 0.19 &  -- & 21.784707 & 8664.8922 & 16:57:13.20&--4:04:11.75 &Y \\
V27 & SR  & $\it 11.915$ &$\it 10.342$  & 0.68 & -- &  -- & 21.040 & -- & 16:57:15.07&--4:05:52.44 &Y\\
V28 & SR & 11.87 & -- & 0.09 & -- &  -- & 60.483833 & -- & 16:57:10.76&--4:04:43.82 &Y \\
V29 & SR  & $\it11.872$ & -- & 0.20 & -- &  -- & 68.388291 & -- & 16:57:27.38&--4:01:24.74 &Y \\
V30 & SR & $\it 12.447$ & $\it 10.858$ & 0.18 & 0.2 &  -- & 71.667981 & -- & 16:57:07.78&--4:06:05.98 &Y \\
V31 & Var?  & 15.882 & 15.690 & -- & -- &  -- & 0.205066 & -- & 16:57:00.63&--4:04:12.50 &Y\\
V32 & sin  & 17.973 & 17.003 & -- & -- &  -- & 0.848041 & -- & 16:57:26.85&--4:04:31.33 &Y \\
V33 & sin & 17.581 & 17.477 & 0.10 &-- &  -- & 0.93353 & -- & 16:57:21.58&--4:02:18.64 &Y\\
V34 & sin & $\it 16.996$ & $\it 16.029$ & -- & -- &  -- & 3.3391 & -- & 16:57:08.49&--4:05:55.68 &?\\
V35$^a$ & SX Phe & 17.147 & 16.641 &  0.18& -- &  0.055261 & -- & 8964.3531 & -- & -- &Y\\
V36$^a$  & sin & 14.470 & 13.219 & 0.025 & -- &  1.082529 & -- & 8964.3990 & --&--&Y \\
V37$^a$ & sin & 14.692 & 13.428 & 0.02 & -- &  0.190840 & -- & 8964.3965 &--&-- &Y\\
N3 & RRc & 16.460 & 15.807 & 0.37 & 0.25 &  0.294386 & 0.294387 & 8287.9063 & 16:57:22.27&--4:04:59.88 &N \\
\hline
\end{tabular}
\raggedright
\center{\quad 
1. These values are intensity-weighted means, except when in italic font which are magnitude-weighted means.
2. All light curves in Fig. \ref {VariablesA} are phased with these periods. When our data were scanty for a proper determination, the period from RO18 was adopted. These periods should be preferred to the ones found in $\S$ \ref{sec:SXfrequencies}.
3. Membership status: Y=member, N=no-member, ?= no proper motion available
$a$. Newly reported in the present work.}
\end{center}
\end{table*}

\begin{figure*}
\includegraphics[scale=0.90]{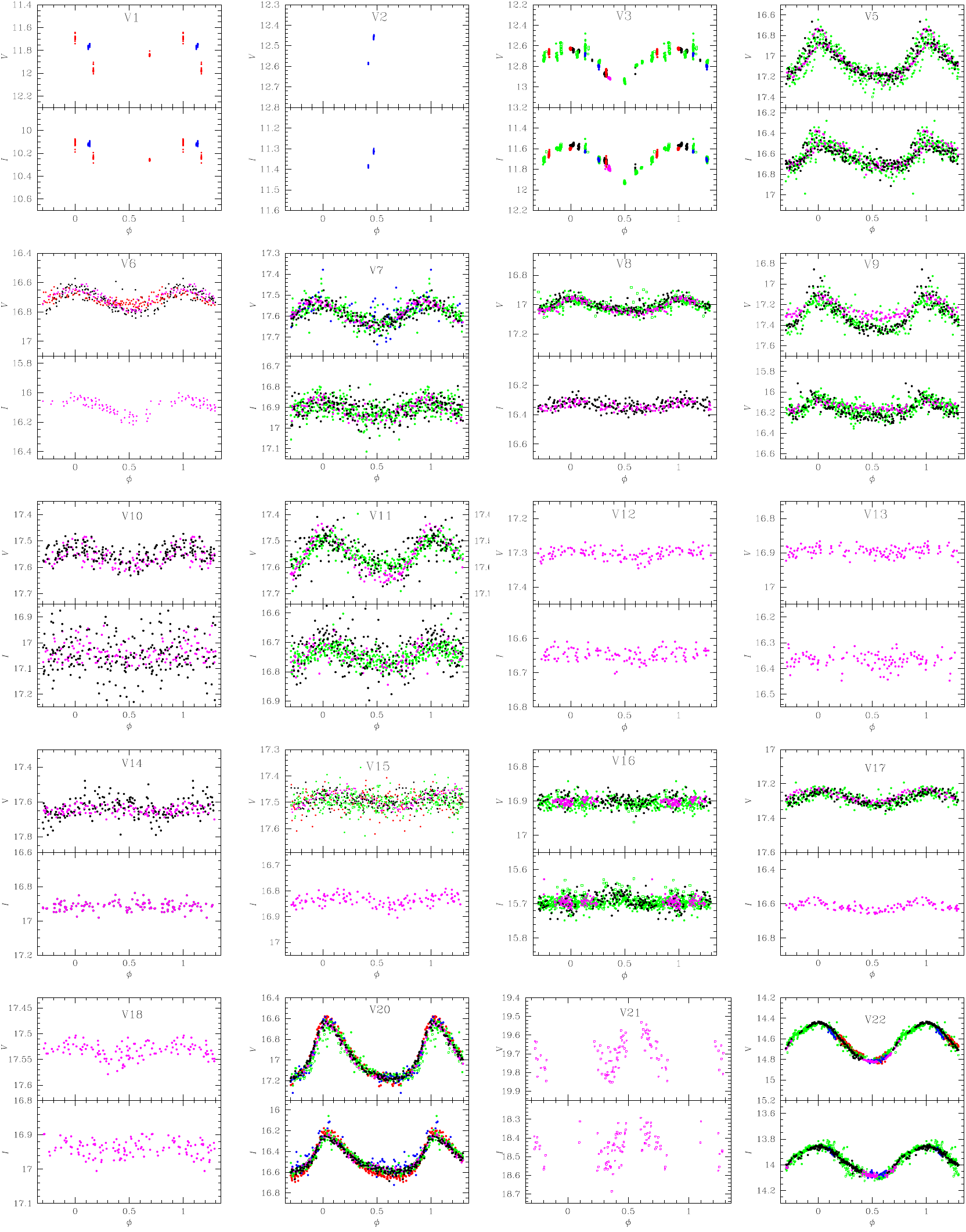}
\caption{Light curves of the variable stars in our FoV phased with the periods listed in Table~\ref{tab:variables}. Black symbols correspond to the run from SPM2018, green from SPM2019, red from BA2018, blue from BA2019 and purple from IAO2020 when available.}
    \label{VariablesA}
\end{figure*}

\begin{figure*} 
%\ContinuedFloat
\includegraphics[scale=0.90]{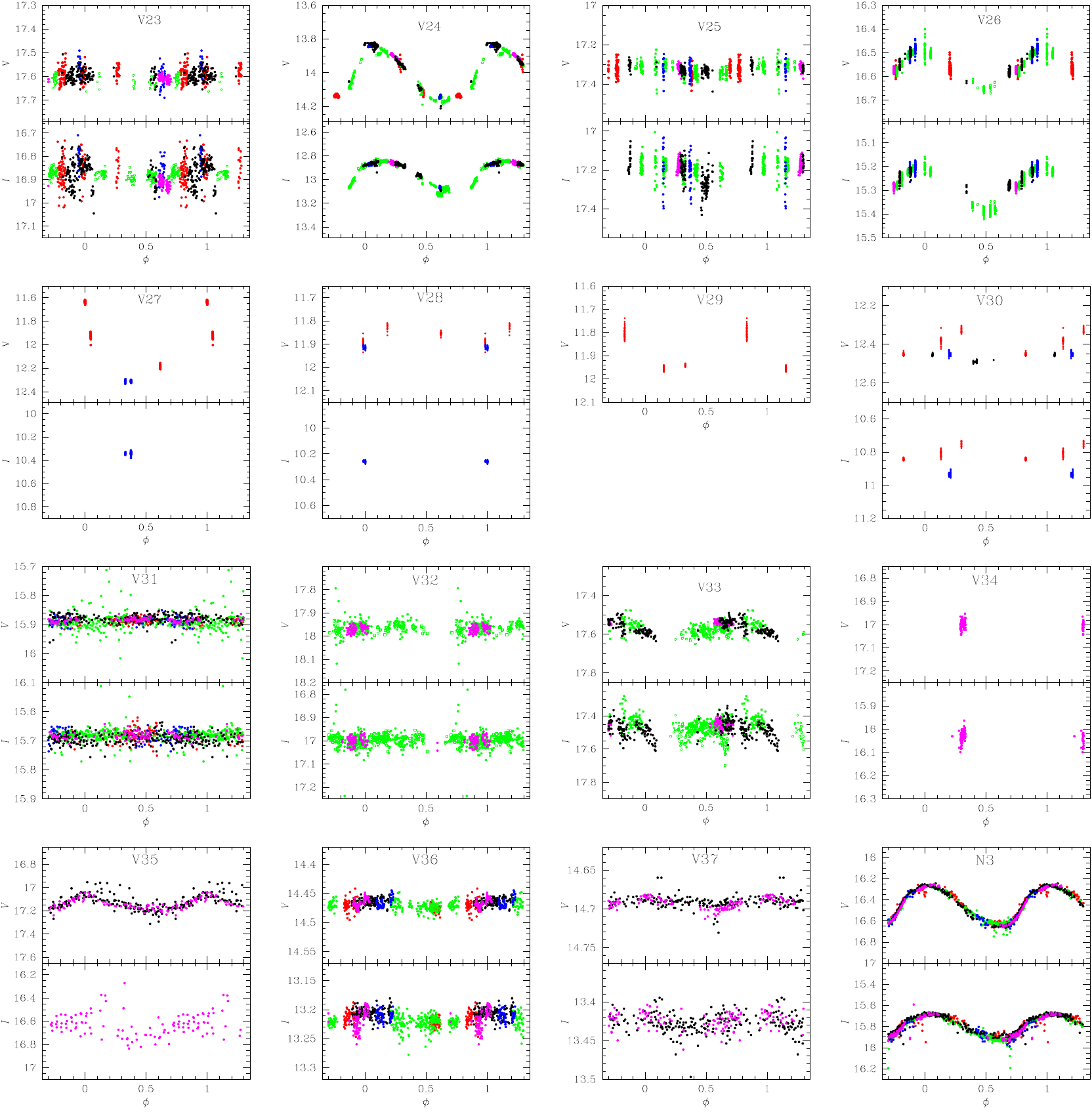}
\addtocounter{figure}{-1}
\caption{Continued}
\end{figure*}

\section{V22: The sole RR\lowercase{c} star in M10, and its physical parameters}
\label{sec:V22}

V22, a RRc star, is the only known RR Lyrae star in M10. RO18 discovered another RRc, labeled N3, but have convincingly argued against the star pertaining to the cluster. The light curve of V22 can be seen in Fig. \ref{VariablesA}. We have Fourier decomposed it and applied the calibrations for RRc stars of \citet{Nemec2013} for the calculation of [Fe/H] and of \citet{Kovacs1998} for the absolute magnitude $M_v$. For all purposes we have adopted  the reddening of the cluster as $E(B-V)=0.25$ obtained from the extinction calibration of \citet{Schlafly2011}, equivalent to
$E(V-I)=1.278 E(B-V)$ \citep{Schlegel1998} or $E(V-I)= 0.319$.

This procedure has been amply discussed in several of our previous papers and for briefness we refer the interested reader in a more detailed description, to the paper by  \citet{Yepez2020}.

In Table \ref{tab:fourier_coeffs}, the Fourier coefficients and the resulting physical parameters, are given. The iron abundance in the spectroscopic scale \citep{Carretta2009},  [Fe/H]$_{\rm C09}$=--1.52 and a distance of 4.67 kpc were found. These numbers can be compared with those reported by \citet{Harris1996} (--1.56, 4.4 kpc). 
We shall discuss these results in the perspective of cluster distance and V22 evolution and membership to the cluster in the following sections.

\begin{table*}
%\scriptsize
\begin{center}
\caption{Fourier coefficients and physical parameters of the RR\lowercase{c} star V22.}       
\label{tab:fourier_coeffs}   
\begin{tabular}{llllllll}
\hline
 $A_{0}$    & $A_{1}$   & $A_{2}$   & $A_{3}$   & $A_{4}$   &$\phi_{21}$ & $\phi_{31}$ & $\phi_{41}$ \\
  ($V$ mag)    & ($V$ mag)  & ($V$ mag)  &  ($V$ mag) & ($V$ mag)& & &  \\
  
\hline
\vspace{0.2cm}
14.637(1)& 0.178(1) &0.006(1) &0.008(1)& 0.006(1)& 4.518(191)&4.771(142)& 2.288(213)\\
\hline
\vspace{0.2cm}

[Fe/H]$_{ZW}$ & [Fe/H]$_{\rm C09}$ &$M_V$ & log~$T_{\rm eff}$  &log$(L/{\rm
L_{\odot}})$ &$D$ (kpc)& 
$M/{\rm M_{\odot}}$&$R/{\rm R_{\odot}}$\\
\hline
-1.59(23)&-1.52(19)&0.516(7)&3.850(1)&1.693(3)& 4.67(2)& 0.48(1)& 4.70(1)\\
\hline
\end{tabular}
\center{Note. The numbers in parentheses indicate
the uncertainty on the last decimal place.}
\end{center}
\end{table*}

\section{The SX Phe stars multimode identification}
\label{sec:SXfrequencies}

All the SX Phe stars in M10 display amplitude and occasional small seasonal phase modulations in their light curves. This is due to the presence of more than one active pulsation mode. Numerous frequencies were listed by RO18
(their table 2)
for the majority of the SX Phe cluster population. In this section we proceed to identify the active frequencies present in our data and their corresponding pulsation mode. For this purpose we employed the algorithms in {\tt period04} \citep{Lenz2005}, where most prominent frequencies (the ones with largest amplitudes), are identified after subsequent pre-whitening of the previous signals.
Given the accuracy, time distribution and number of independent measurements in our data collection, the number of frequencies that can be confidently identified is limited, even if weaker frequencies are actually present. We stopped the pre-whitening process once the amplitude of the remaining signal is lower than 10 mmag, as we do not feel we can trust the identifications of weaker signals. In all cases this enables to isolate 2-3 frequencies.

In Table \ref{tab:SXmodes} we list the frequencies, and their corresponding periods, of the detected signals in the frequency spectrum, in order of decreasing amplitude.
Exploring the period ratios, we can identify in several cases, the fundamental ($P_0$) and first overtone ($P_1$) or suspect the presence of non-radial modes. For this purpose it is convenient to recall that typical period ratios in SX Phe stars are $P_1/P_0 = 0.783$ and $P_2/P_0 = 0.571$ \citep{Jeon2003,Jeon2004}. Combining this approach with the position of a given star in the P-L plane in Fig. \ref{SX_PL}, we assigned the pulsation mode in column 6 of Table \ref{tab:SXmodes} whenever it was possible. We note that the main or primary frequency detected by {\tt period04} agrees well with the period found via the string-length method, given in Table \ref{tab:variables} which was used to phase the light curves in Fig. \ref{VariablesA}.
Although some of the SX Phe variables show signals with low frequencies
(f < 10d$^{-1}$), these are probably artificial, and are caused by small
zero-point differences between the data sets.
We retain them in the list as they appear but we do not try to associate them to a pulsation mode.

\begin{table*}
\footnotesize
\begin{center}
\caption{Pulsation modes in the SX Phe stars}
\label{tab:SXmodes}
\begin{tabular}{lcccccc}
\hline
Star  & id. &Freq. & $P^1$ & Amp.&mode$^{2}$ &comment\\
\hline
& &d$^{-1}$ & d &mmag.& &\\
\hline
 V5 & $f_1$ &17.079488 & 0.058550 & 175 & $P_0$   & $P_1/P_0=0.746$\\
    &$f_2$&34.161476& 0.029273 &56& 2$P_0$&\\
  &$f_3$& 22.881312&0.043704 &39&  $P_1$&\\
   &$f_4$& 3.769797&0.265266 &34 &&\\
   &$f_5$& 6.919094&0.144528&22& &\\
V6 & $f_1$ &15.692486 & 0.063724 & 55 & $P_1$   & $P_1/P_0=0.835$?\\
&$f_2$&13.100973& 0.07633 &28& $P_0$&\\
  &$f_3$& 21.314535&0.046916 &17&  &\\
   &$f_4$&2.2062028&0.453267 &11 &&\\
   
 V7 &$f_1$ &20.787514& 0.048106&48 &$P_0$&\\
 &$f_2$ & 20.347529&0.049146&  14 &nr&\\
 
 V8&$f_1$ & 19.6051124& 0.051007&34& $P_1$ &$P_1/P_0=0.720$  \\
 &$f_2$ &  0.966971&1.034158& 13& & \\
 &$f_3$ &  39.175446& 0.025014&10 &  2$P_1$&  \\
  &$f_4$ &  14.122000& 0.070812&9& $P_0$ &\\ 
  
V9&$f_1$ &19.488544& 0.051323&127&$P_0$&  \\
  &$f_2$ & 2.507949& 0.398732& 47&     &  \\
  &$f_3$ &38.977162& 0.025656& 46&2$P_0$&  \\
  &$f_4$ &18.985092& 0.052673& 26&  nr  &  \\
  
V10&$f_1$ & 45.307503&0.022071& 25& $P_2$&  \\
   &$f_2$ & 39.141519&0.025548& 20& ?   &  \\      
   &$f_3$ & 3.2389789&0.308739&12&&  \\
   &$f_4$ & 31.798354&0.031448&9& $P_1$ &$P_2/P_1=0.702$  \\ 
   
V11&$f_1$ & 20.851572&0.047958& 61& $P_0$ & \\
   &$f_2$ & 20.872999&0.047909& 17&  nr? & \\
   &$f_3$ & 41.927681&0.023851& 13&2$P_0$ & \\
   &$f_4$ & 1.7971021&0.556451& 12& & \\

V12&$f_1$ & 42.816506&0.023355 &11&  & \\

V13&$f_1$ & 5.059147&0.197662&10& $\sim$  3$P_0$  & \\
   &$f_2$ &15.851994&0.063084&9&$ P_0$  & \\

V14&$f_1$ &23.366096&0.042798 & 30&$P_0$&$ P_1/P_0=0.821$\\ 
   &$f_2$ &28.473315&0.035121 & 16&$P_1$ &  \\   
   &$f_3$ &21.559168&0.046384 & 15&   & \\

V15&$f_1$ & 28.706127&0.034836 &22&$P_1$& \\
   &$f_2$ & 25.912539&0.038591 &11& $P_0$?& \\
   &$f_3$ & 35.179626&0.028426 & 9& $P_2$& $P_2/P_1$=0.816?\\

V17&$f_1$ &27.066621&0.036946 & 35 &$P_1$&\\
   &$f_2$ & 1.913102&0.522711 & 10 &&\\
   &$f_3$ &27.203234&0.036760 & 9 &&\\

V18&$f_1$ & 25.633011&0.039012&12  & $P_0$&\\ 

V20&$f_1$ & 19.761553& 0.050603&252&$P_1$ &\\
   &$f_2$ & 39.523185&0.025302 & 80&$P_1$/2 &\\
   &$f_3$ & 59.284886&0.016868 & 26&$P_1$/3 &\\
   &$f_4$ & 18.798335&0.053196 & 23& & \\
   
V35&$f_1$ &17.875652&0.055942& 53& $P_0$&$P_2/P_0$=0.478\\
   &$f_2$ & 7.082806&0.141187& 19& & \\
   &$f_3$ &37.420082&0.026724& 17& $P_2$?& \\  
\hline
\end{tabular}
\center{1. These periods generally agree well with those from the string-length method listed in column 7 of Table \ref{tab:variables}. However some differences between 1-5\% are seen in a few stars. Periods in Table \ref{tab:variables} shall be preferred as they phases better the present light curves. 2.$P_0$: Fundamental radial mode; $P_1$: First overtone radial mode; $P_2$: Second overtone radial mode; nr= likely non-radial mode.}
\end{center}
\end{table*}

\section{On the distance to M10 from independent methods}
\label{Sec:distance}

\subsection{The SX Phe stars P-L relation}
\label{Sec:SXPHE}

M10 is a globular cluster with a rich population of SX Phe stars, most of them cluster members. Their position in the Blue Stragglers region of the CMD can be seen in Fig. \ref{CMD}

These SX Phe will play a role in the determination of the distance via their well known period-luminosity (P-L) relation. We have considered three independent calibrations of the P-L  relationship; \citet{Poretti2008}, \citet{Arellano2011} and \citet{CohenSara2012} given explicitly in the
eqs. \ref{PLP}, \ref{PLAF} and \ref{PLCS}, repectivelly

\begin{equation}
\label{PLP}
M_V = -3.65\, \log P - 1.83.
\end{equation}

\begin{equation}
\label{PLAF}
M_V = -2.916\, \log P - 0.898.
\end{equation}

\begin{equation}
\label{PLCS}
M_V = -3.389\, \log P -1.640.
\end{equation}

\begin{figure}
\includegraphics[scale=0.43]{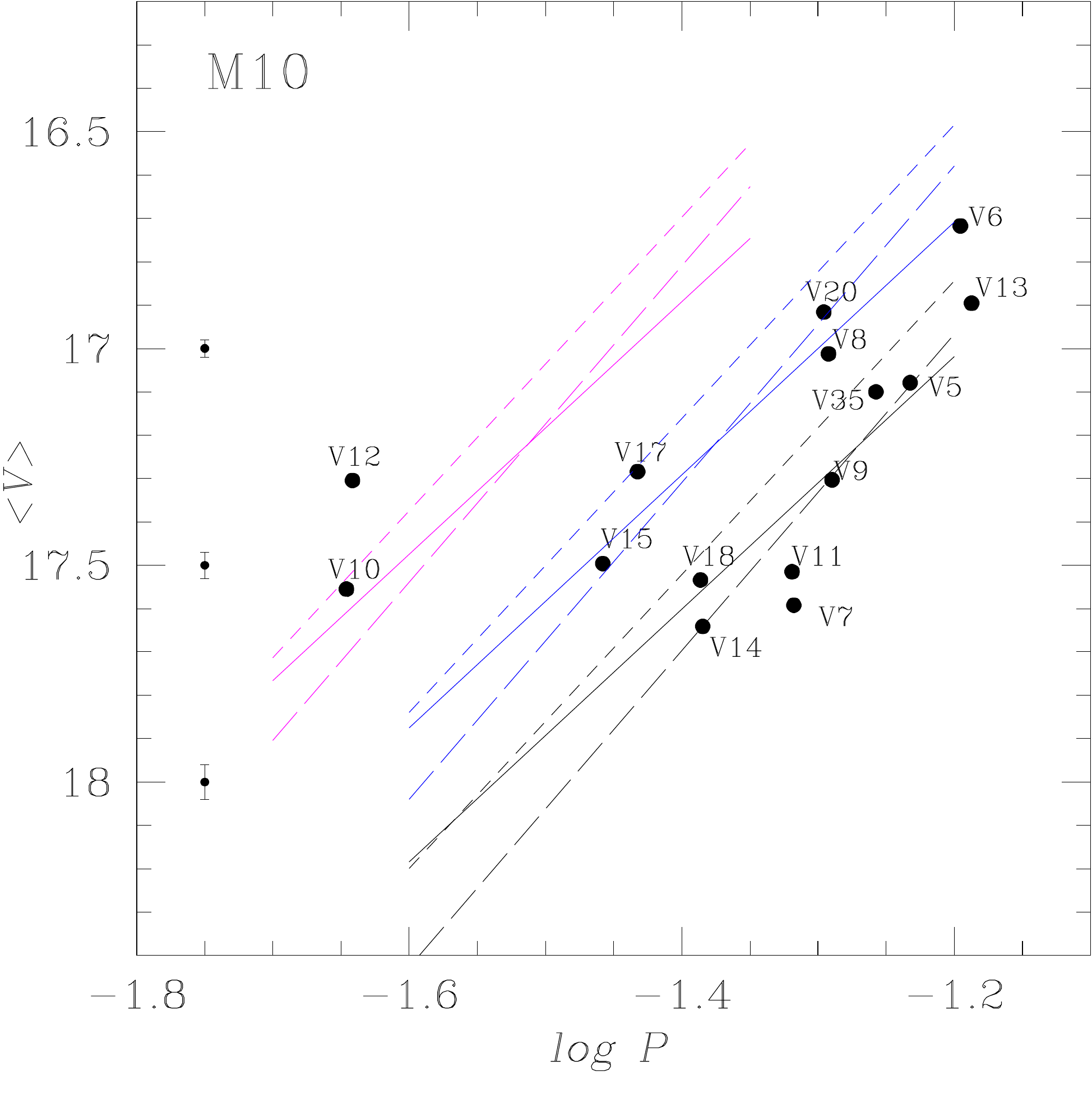}
\caption{The P-L relationship for SX Phe stars. Three calibrations are shown;\citet{Poretti2008} (long-dash), \citet{Arellano2011} (solid) and \citet{CohenSara2012} (short-dash), and for fundamental (black), first overtone (blue) and second overtone (magenta). The calibrations have been shifted to a distance of 5.35 kpc and $E(B-V)=0.25$. (see $\S$ \ref{Sec:SXPHE} for a discussion).}
    \label{SX_PL}
\end{figure}

The period $P$ in the above equations corresponds to the radial fundamental mode. In Fig. \ref{SX_PL} we display the log~$P$ - <V> plane with the distribution of 14 of the 15 known SX Phe stars in M10. V19 is out of the FoV of our images. To calculate the cluster distance from the SX Phe via the above relations, we proceeded as follows; to identify the pulsation mode of the principal period in the power spectrum of each star, we overlaid the P-L relationships shifted to the distance suggested by the ZAHB level
seen in the CMD of Fig. \ref{CMD}, i.e. 5.19 kpc, and assuming $E(B-V)=0.25$. For the first and second overtone relations (blue and magenta lines in the figure), we adopted the period rates $P_1/P_0 = 0.783$ and $P_2/P_0 =
0.571$ (see Santolamazza et al. 2001 or Jeon et al. 2003;  Poretti et al. 2005). These adopted values are typical period ratios.
 The dominant mode was identified as the fundamental mode for V5, V7, V9, V11, V13, V14
and V18; as the first overtone for V6, V8, V15, V17 and V20; and as the second overtone
for V10 and V12. Eqs. \ref{PLP}, \ref{PLAF} and \ref{PLCS} were applied to these modes (taking into account the
appropriate period ratios for the overtones), and these results were averaged for each of
the stars.
The overall mean, in which we ignored V9 due to its peculiar position in the CMD, and V12 for being too bright for this given period, is 5.36 $\pm$ 0.32 kpc. Fig. \ref{SX_PL} shows the P-L relations shifted to this distance, as detailed in the caption.
 
It shall be noted that the distance obtained for the RRc star V22, via the Fourier light curve decomposition is some 600 pc shorter than the SX Phe result from the P-L relation. In fact, as noticed before, V22 lies about 0.2-0.3 mag above the ZAHB. In our analysis of the $Gaia$-DR2 proper motions ($\S$ \ref{membership}), V22 was encountered to be a likely cluster member, thus, its higher luminosity may be a consequence of some advanced evolution post-ZAHB.
 
 \subsection{The RR Lyrae P-L ({\it I}) relation.}

 Another approach to the distance determination via RR Lyrae stars is using the P-L ($I$) calibration for RR Lyrae stars of \citet{Catelan2004}, $M_I = 0.471-1.132~ {\rm log}~P +0.205~ {\rm log}~Z$, with ${\rm log}~Z = [M/H]-1.765$; $[M/H] = \rm{[Fe/H]} - \rm {log} (0.638~f + 0.362)$ and log~f = [$\alpha$/Fe], from where we adopted [$\alpha$/Fe]=+0.4 \citep{Sal93}. Given the period 0.404485 d, we found 
a distance of 4.9 kpc. This places the cluster a bit further than expected.

\subsection{The tip of the RGB}

Yet another approach to estimate the cluster distance is using the variables
near the tip of the RGB (TRGB).
This method, originally developed to estimate distances to nearby galaxies (Lee et
al. 1993) has already
been applied by our group for the distance estimates of other clusters e.g. \citet{Arellano2015}  for
NGC 6229 and \citet{Arellano2016} for M5. In the former case the method
was described in detiled.  In brief, the idea is to use the bolometric magnitude
of the tip of the RGB as an indicator. We use the calibration of \citet{SalarisCass1997}:

\begin{equation}
\label{TRGB}
M_{bol}^{tip} = -3.949\, -0.178\, [M/H] + 0.008\, [M/H]^2.
\end{equation}
 
However one should take into account the fact that the true
TRGB might
be a bit brighter than the brightest observed stars, as argued by \citet{Viaux2013} 
in their analysis of M5, under the arguments that the neutrino magnetic dipole moment
enhances the plasma decay process, postpones helium ignition in low-mass stars, and
therefore extends the red giant branch (RGB) in globular clusters. According to these
authors the TRGB is between 0.05 and 0.16 mag brighter than the brightest stars on the RGB. The brightest member star in M10 is the star labeled V27 in the CMD,
and applying the corrections 0.05 and 0.16, we find that the distance to the cluster must be between  5.1 kpc and
4.8 kpc.

In Table \ref{tab:distance} a summary of the distance values found by the above methods is presented.

\begin{table}
\footnotesize
\caption{Summary of M10 distance estimates from several methods. }
\centering
\begin{tabular}{ll}
\hline
$D$  &  method \\
 kpc     &  \\
\hline
 4.7   & Fourier decomposition for RRc star V22\\
 5.3$\pm$0.1   & Positioning of isochrones on CMD\\
5.3$\pm$0.3& SX Phe P-L relation for member stars\\ 
4.9& P-L(I) relation for RR Lyrae stars\\
4.8-5.1&TRGB  \\
\hline
\end{tabular}
\label{tab:distance}
\end{table}

\section{The Colour-Magnitude diagram}
\label{sec:CMD}

\begin{figure*}
\begin{center}
\includegraphics[scale=1.0]{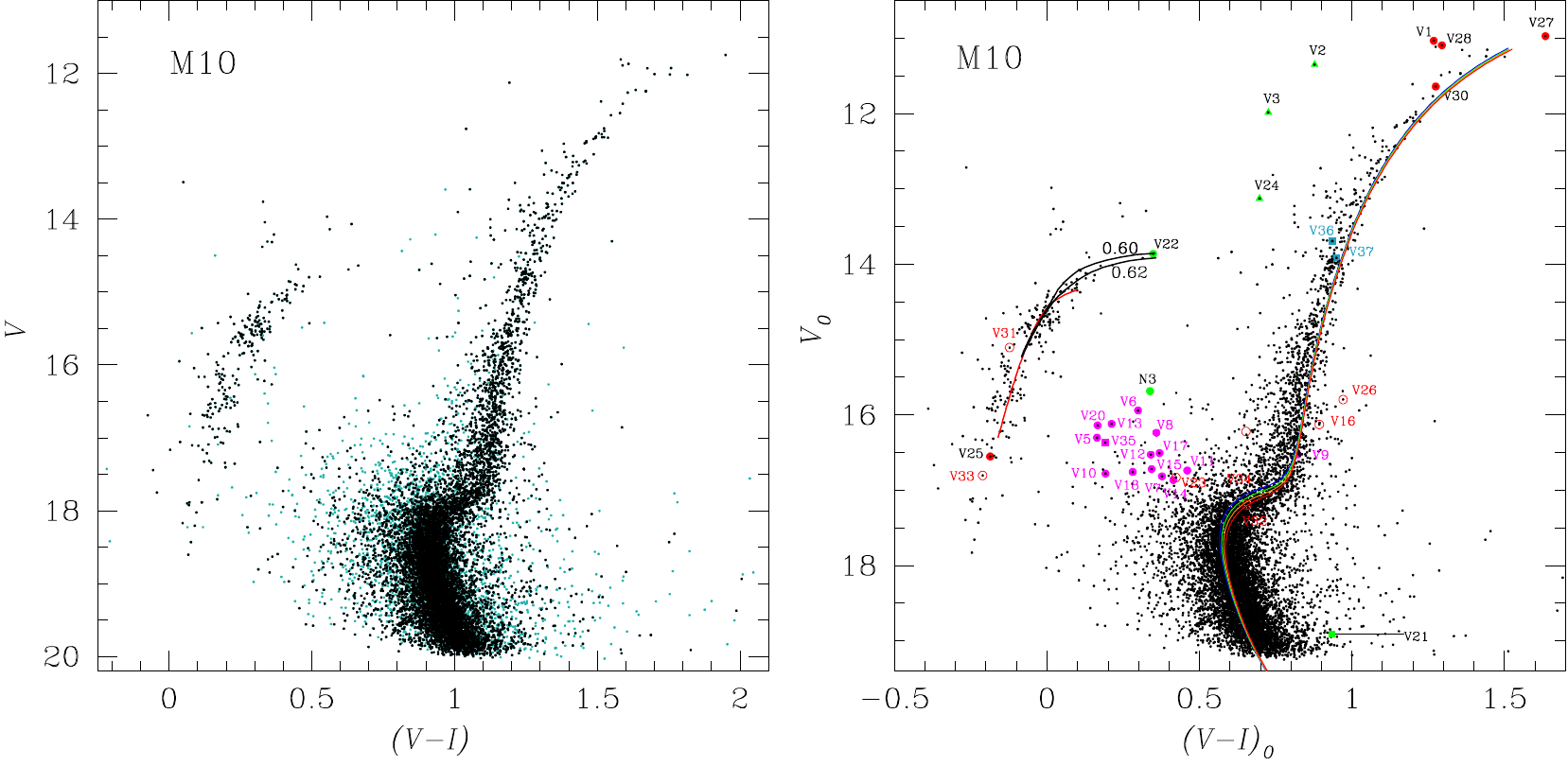}
\caption {The CMD of M10. In the left panel the member and non member stars are shown in black and light blue respectively. The right panel shows the deredenned CMD of member stars, for which $E(B-V)=0.25$ was adopted. The colour code for variable stars is: solid green, purple and red circles represent RRc, SX Phe and SR variable stars respectively. Green triangles and open red circles are used for W Vir/BL Her and unclassified variables respectively. The blue tail portion of the ZAHB and isochrones are calculated from the models of \citet{Vandenberg2014} for [Fe/H]=$-1.58$, Y=0.25, [$\alpha$/Fe]=+0.4, and ages 12.0, 12.5, 13.0 and 13.5 Gyr. The black loci show the evolutive tracks for a 0.60 and 0.62 $M_{\odot}$ described in $\S$ \ref{HBmodels}. All theoretical loci were placed for a cluster distance of 5.35 kpc.}
\label{CMD}
\end{center}
\end{figure*}

The CMD of the cluster is shown in Fig.~\ref{CMD}. In the left panel, member and non-members found by the analysis of $\S$ \ref{membership} are shown. In the right panel, the dereddened version of the CMD is presented and the location of all the
known variables in our photometry is marked. All variable stars are plotted using their intensity-weighted
means $< V >_0$ and corresponding colour $<V>_0 - <I>_0$.  Most variables have a clear counterpart in the $Gaia$-DR2 data base with a proper motion measurement. Their membership status is indicated in the last column of Table \ref{tab:variables}.
For star without a proper motion (e.g. V8, V9, V14, V25, V34, V35) we assigned their status based on the combination of its variable type and position on the CMD, and other considerations, e.g. V35, a SX Phe star that follows the P-L relationship discussed in $\S$ \ref{Sec:SXPHE}.

In the diagram we include the blue tail of the zero-age horizontal branch (ZAHB) and some isochrones calculated from the models of \citet{Vandenberg2014} for [Fe/H]=$-1.58$, Y=0.25 and [$\alpha$/Fe]=+0.4, for an age range of 12.0-13.5 Gyr. Also shown are two evolutionary tracks from the hot region of the HB which will be discussed later in $\S$ \ref{HBmodels}. We shifted all the stars by $E(V-I) = 0.319$ and different distances, searching the best match with the theoretical sequences. We found a proper placing for $d=5.35$ kpc, the distance suggested by the SX Phe star P-L relation. We note that the distance for the RRc V22, found from the Fourier decomposition, 4.67 kpc is significantly shorter, and that the star lies $\sim$ 0.2-0.3 mag above the ZAHB.

Without pretending to perform a cluster age determination, we note that the member stars distribution on the CMD and the older isochrones of 13-13.5 Gyrs are more consistent, suggesting that the cluster is older than the 11.75$\pm$0.38 Gyrs estimated by \citet{Vandenberg2013}

\section{Modelling the Horizontal Branch of M10}
\label{HBmodels}

\begin{figure}
\includegraphics[scale=0.45]{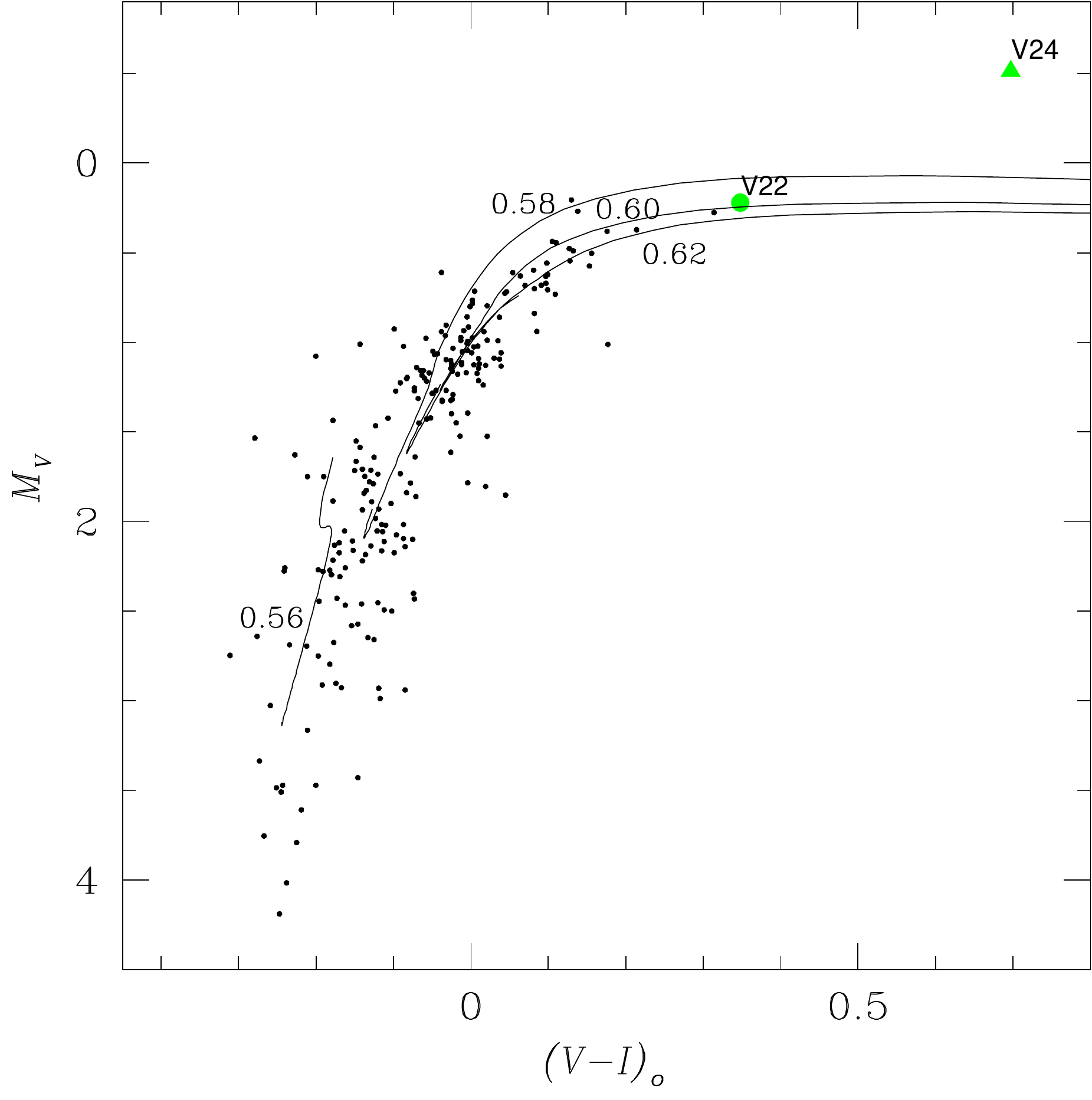}
\caption{The well populated HB of M10 allows us to model it well, (see $\S$ \ref{HBmodels} for details).
A mass range of 0.56 to 0.62 $M_{\odot}$ is needed to reproduce well its full length down to the blue end. Common progenitor of these post-He-flash
models is a 13.0 Gyr old model of an initial mass of 0.83 $M_{\odot}$.}
    \label{evolution}
\end{figure}

Given the here presented, good photometry and critical membership
assessment of the rich stellar content of M10, its HB is very well
defined in the $M_v$ versus $(V -I)_0$ diagram of Fig. \ref{evolution} (see also Fig. \ref{CMD}). 
That precise and ample observational evidence allows us 
to model the mass and age of M10 HB stars with better accuracy, 
than it is possible for main sequence stars near the very broad and 
slow turn-off point (Fig \ref{CMD}).

For this purpose, HB star colours are not as critical as they are
for modelling stars near the turn-off, where we used the
well-tested VandenBerg models, which also have a fine metallicity grid.
Rather, for an understanding of the formation of HB stars, the history
of RGB mass-loss matters a lot. To address this issue suitably,
we here now use our own evolution models, as we will further explain 
below.

For over half a century now, we see a discussion of the details of 
the formation history of the HB, and how to understand the large 
cluster-to-cluster variety of its population. First of all, there are 
large differences in how far towards the blue a HB is populated. M10 
here is quite an extreme case, showing no yellow HB stars, while other 
globular clusters have also many yellow and white, but no blue HB stars. 

From an empirical point of view (since \citet{Sandage1960}, and see \citet{Gratton2010}
for a brief, nice review), there is a clear relation 
of this "first parameter" with metal content: The lower the metallicity, 
the bluer is the HB stellar population. But from a theoretical point 
of view, the effective temperature of a HB star depends on two
factors: (i) the metallicity, by virtue of lesser opacities, allowing for 
a more compact shell with a hotter and bluer photosphere, 
and (ii) the mass of the Hydrogen-rich 
shell: Comparing models of the same metallicity but different shell 
masses show (see below), that a lower shell mass also leads to a bluer 
HB star. This fact was already pointed out and exploited by \citet{KPS2005}, (hereafter referred to as SC2005). With shell masses 
reaching 0.3 $M_{\odot}$, there is then no substantial difference left, when compared to a normal K giant
clump star in central Helium-burning, because the observable properties of such central Helium-burning stars
only change marginally with further increase in the shell mass.

Obviously, HB stars in older globular clusters have developed from slightly 
lower mass stars on the main sequence, compared with younger globulars,  
and since the degenerate Helium-core needs in all cases 0.5 $M_{\odot}$ 
(within a narrow margin) to start central Helium-burning -- by means of 
the "Helium flash" on the tip RGB -- the resulting HB stars then have a 
smaller Hydrogen-rich shell mass, and, consequently, are
bluer. At the same time, such older globular clusters tend to be less 
metal-rich (however, there are notable exceptions, apparently having 
formed close to the metal-richer, early Milky Way centre). For this
tendency, which then adds to the above mentioned physical dependence on 
metallicity, the physical dependence on the shell-mass of the HB stellar 
colours is entirely veiled by the empirically found metallicity 
dependence. 

In addition, there remains debate about why the stretch of the HB population 
covers a larger range of effective temperature (or colour), than one 
should expect from a single suitable model of such a HB star and its moves 
in the HRD. This phenomenon, dubbed the "second parameter problem",
was brought up by \citet{VandenBerg1990}, 
who noticed that the stretch of many HBs is wider than what could be 
explained by a reasonable variety in age of the cluster populations.

Consequently, this observation leaves only one other simple explanation: 
A slight variation of the mass loss on the RGB, which results in a
certain range of shell mass of HB stars of the same globular cluster,
despite their initial masses and ages all being almost identical.
In the following we are going to prove, that by this simple idea, 
the stretch of the HB population of M10 can already be fully 
reproduced.

When discussing the respective RGB stellar mass loss, again
metallicity seems to play a role, assuming an empirical point of view,
see especially \citet{McDonald2015}. Using the simple Reimers 
mass-loss law with one and the same constant ($\eta$) does not
produce the right variation of shell masses needed to reproduce the
different HB populations found in globular clusters of different 
metallicity.

However, as already pointed out by SC2005, there is no need for any 
such empirical dependence of $\eta$ on metallicity, if only a physical 
approach is taken.
When the resulting modification of the Reimers law by two extra terms
is then applied to the RGB mass-loss, such evolution models reproduce the
HB shell masses needed for very different metallicities with one and the
same value of $\eta$. This is mainly due to the fact that one of these 
additional terms depends on effective temperature. Since metal-poor
globular cluster RGB stars are bluer, this physically motivated term 
produces essentially the same effect as does an empirically motivated 
metallicity term. 

The main point of SC2005 is to consider the chromospheric mechanical flux
to be the energy reservoir, which provides the cool, Reimers-type 
stellar wind, which applies to RGB stars, because radiation pressure 
on dust is not a relevant factor here. The other additional factor is
gravity-dependent and derives from the growing (with lower gravity) 
extent of the self-sustained chromosphere along the RGB. SC2005
calibrated it with the chromospheric properties of the well-studied 
and pivotal K-supergiant zeta Aurigae.

We here used the same evolution code and parameterization, especially 
the same prescription of the RGB mass-loss by that modified Reimers
law with with $\eta$ = $0.8 \cdot 10^{-13}$, as presented by SC2005. 
This evolution code was originally
developed by Peter Eggleton
\citep[see][]{Eggleton1971, Eggleton1972, Eggleton1973}
 and further improved and thoroughly 
tested by \citet[][]{Pols1997, Pols1998}  and \citet{KPS1997}. We here use abundances for Z=0.001, 
equivalent to [Fe/H]=–1.3, close enough to the metallicity found
for M10.

To compare models and cluster photometry in the HRD (see Fig. \ref{evolution}), 
we dereddened the observed $V-I$ colours by $E(V-I)$ = 1.278 $E(B-V)$= 0.319,
and find that the distance modulus matches best the absolute visual
brightness of the HB models on the well-defined right-hand, 
horizontal part of the HB for a distance of 5.35 kpc, as such 
confirming the result derived above from the SX Phe type variables.

As a result, the blue end of the M10 HB coincides well -- within some
scatter (see Figs. \ref{CMD} and \ref{evolution}) caused by the photometric uncertainty, -- with
our evolution model of HB stars with a total mass of 0.56 $M_{\odot}$,
a Helium core mass of 0.50 $M_{\odot}$, and a Hydrogen-rich shell 
mass of only 0.06 $M_{\odot}$. Such stars spend a very short fraction 
of their life-time on the HB, only $\sim$ 100 Myrs. Consequently, relatively 
few stars of any such cluster are, at any given time, in this state,  
and a massive globular cluster like M10 is required to get to study 
sufficiently large numbers. 

These HB stars, which match the M10 HB population, descend from a 
pre-He-flash evolution model with an initial mass of 0.83 $M_{\odot}$ 
(on the zero age main sequence), and therefore have an age of 13.0 Gyrs. 
This relatively large age is in very good agreement with the isochrone 
age analysis above, based on the models of VandenBerg et al. (2014),
and contributes indirectly to the far blue-stretched HB, as with older
age more stellar shell mass is lost on the upper RGB (see below).
We find from our evolution models, that most of the mass needed to be 
lost to reduce the HB shell mass sufficiently (i.e. 0.27 $M_{\odot}$), 
is actually lost on the upper RGB, before finally the fully degenerate
Helium core ignites.

During the central Helium burning phase, our 0.56 $M_{\odot}$ HB
evolution model only covers the lower-left part of the observed
HB stellar population (see Fig. \ref{evolution}). Evolution then accelerates and
this model alone cannot, therefore, explain the even denser stellar
population to the upper right of the HB. Assuming a uniform age
of the globular cluster population, this suggests a simultaneous
presence of HB stars, which have somewhat larger shell masses,
reaching up to a total mass of 0.62 $M_{\odot}$. These stars must have
lost up to 30\% less mass on their upper RGB evolution than the
precursor of the 0.56 $M_{\odot}$ HB evolution model.

Together, the slower evolving, central Helium-burning stages of 
these models cover exactly the observed HB population 
of M10, when allowing for some scatter due to observational errors. 
We should add the note, that this approach is different from
comparing with a zero-age HB isochrone, because then the evolved
and slightly brighter and bluer stages of central Helium-burning
are not taken into consideration.     

A moderate star-to-star variation of the mass-loss on the RGB would 
therefore make a simple (no other parameters involved) and natural 
explanation of the "second parameter problem". But what would then 
be a plausible cause for such a variation? In which way can stars of 
the same age and mass in a globular cluster differ to not all of 
them reach the same mass loss? --
Since some time now, we know of the presence of a varying degree 
of magnetic field in red giants, \citep[see, e.g.][]{Konstantinova2013}. We may therefore speculate, that the corresponding,
individually different coverage by closed magnetic field can hinder
a fraction of the prescribed mass-loss (apparently up to 30\%). 
– The rich and old globular M10 provides a perfect
testing ground for this question.

\section{Summary of results}
\label{sec:Summ}

We have performed a new CCD photometric study of the globular cluster M10 based on CCD images obtained in three sites during 22 nights in 2018, 2019 and 2020.
We have analyzed the variable stars individually with the aim to
confirm their identification,  classification and membership in the cluster.

The corresponding equatorial  coordinates for all the variables in the FoV were either confirmed or corrected and a detailed finding chart is offered, which is most useful particularly for the faint variables in crowded regions. A search in our light curves collection lead to the discovery of a new SX Phe star, likely a cluster member (V35) and two sinusoidal variables, also cluster members in the RGB, whose classification remains unclear (V36 and V37).

Fourier decomposition of the light curve of the only RR Lyrae known in the cluster, V22, lead to the estimation of the iron abundance in the spectroscopic scale [Fe/H]$_{spec}=-1.52\pm 0.19$ (or [Fe/H]$_{ZW}=-1.59\pm 0.23$) , consistent with high resolution spectroscopic determinations, e.g. =-1.52$\pm$0.02 \citep{Kraft1995} or the mean adopted by \citet{Harris1996} of -1.56.
The distance estimation from this approach 4.67 kpc, is too small compared with the results of other independent methods. V22 is about 0.2-0.3 mag above the ZAHB; being a likely cluster member from its proper motion analysis, 
the reason of its high luminosity may be, at least partially, due to evolution off the ZAHB.
The distance estimated via the positioning of isochrones on CMD; the SX Phe P-L relation for member stars; the P-L(I) relation for RR Lyrae stars and the estimation from the TRGB, ranges between 4.9 and 5.3 kpc. These distances are consistently larger than the standard distance accepted of 4.4 kpc \citep{Harris1996}, obtained from the $M_V$-log $P$ calibration.

The $Gaia$-DR2 proper motions of the 9249 stars  the FoV of our cluster images, for which we posses light curves, and the approach of \citet{Bustos2019} enable a cleaner version of the CMD. The isochrones from \citet{Vandenberg2014} and the distribution of member stars near the turn-off point, suggest a cluster age of $\sim$ 13 Gyrs.

The SX Phe stars are the best represented variable population in M10, however, of the 16 known SX Phe in the FoV, four of them are likely non cluster members, namely V8, V9, V11 and V14. 

Virtually all the SX Phe light curves display amplitude and phase modulations, clearly due to the presence of multiple mode pulsation. A successive prewhitening process of the frequency spectra allowed identifying 2-3 active periodicities, which combined with the distribution of stars on the P-L plane and three independent P-L calibrations, suggest an identification of the pulsation mode.

Other variables that were found likely non cluster members are V25, V34. The other RRc star in the FoV, N3,
is definitively not member of the cluster but a more distant star.

The blue tail of the HB was modelled using the evolution code and parametrization, particularly 
 the RGB mass-loss by a modified Reimers law (SC2005). The resulting distance 5.35, matches the distance found from the SX Phe variables P-L relation. It was found that a 0.83 $M_\odot$ model at the main sequence, lost some 30\% of its mass at the upper RGB. The remaining core He-burning star of 0.56 $M_{\odot}$, descends to the blue HB, completing its MS-HB journey in about 13 Gyrs.

\vskip 2.0cm

\noindent
AAF acknowledges the support from DGAPA-UNAM grant through project IG100620. MAY thanks CONACyT for the PhD scholarship. We thank the staff of IAO, Hanle
and CREST, Hosakote, for making the observations possible. The authors are grateful to the referee for valuable and constructive comments. The
facilities at IAO and CREST are operated by the Indian Institute
of Astrophysics, Bangalore.
We have made an extensive use of the SIMBAD and ADS services, for which we are
thankful.

{\bf Data Availability:} The data underlying this article shall be available in electronic form in the Centre de Donnés astronomiques de Strasbourg database (CDS), and can also be shared on request to the corresponding author

\bibliographystyle{mnras}
\bibliography{M10_MNRAS} % if your bibtex file is calledexample.bib

\appendix
\label{Appendix}

\section{Comments on individual stars}

In this section we address only those stars whose light curve, classification, identification, membership status or position in the CMD trigger some controversy or deserve a comment.

V8. The cluster membership of this star was doubted by RO18 on the base of the CASE proper motions. We encounter that the best coincident $Gaia$-DR2 source lacks proper motion measurement, therefore we cannot  be conclusive on its membership status.

V9. Striking differences in the colour of this star are evident in the results of SA19, RO18 and the present work, that found the star in the blue tail of the HB, in the BS region and in the RGB respectively. In our opinion these discrepancies are driven by blending issues in all cases.

Due to its large amplitude and brightness, it was suggested by SA16 that the star may be a foreground $\delta$ Scuti.  Discrepancies between the CASE and $Gaia$ proper motions were noticed by RO18, and in fact we confirmed that the 
best $Gaia$-DR2 source coincidence with the star coordinates, lacks of a proper motion measurement, making it impossible for us to accurately assess its membership status. Blinking the residual images in our IAO2020 collection we found that the variability happens slightly to the NE of the otherwise identified star by its coordinates. Our light curve is most likely the result of a blending of the true variable with neighbours, which badly contaminates the colour, if not so much the magnitude.

Its indisputable variability is evidenced by a nice large-amplitude light curve showing modulations. While SA16 did not find secondary frequencies, probably due to the short time-span of 6.7 hours of their data, we encountered a significant period  (see Table \ref{tab:SXmodes}) of 0.052673 d of probably non-radial nature and responsible for the amplitude modulations.

V12. This star displays a light curve extremely low amplitude and noisy in the discovering paper \citep{Salinas2016}, however we detect clear variations in our IAO2020 data, consistent with the period of RO18, $P = 0.022823$ d. However, in the SX Phe P-L relation the star is too bright for its short period, even for a second overtone pulsator. While the star is likely a cluster member, we did not include it in the distance estimation.

V14. The light curve of this star displays variations of very small amplitude in our IAO2020 data, rather consistently with that of \citet{Salinas2016}, obtained from much higher resolution data.
It lies in the blue straggler region, moving to a classification as a SX Phe. We could not find a $Gaia$-DR2 counterpart, hence we cannot pronounce on its cluster membership. Taking it as $bona fide$ member (RO18), in the SX Phe P-L (Fig. \ref{SX_PL}) relation it falls in the fundamental mode calibration loci. However, it was not considered for the distance calculation.

V16. The star is listed as a suspected variable by RO16. We also do not detect variations even in our best quality data. Although the star was found a cluster member, it probably should be considered a constant star.

V21. This star was found to be a contact binary by RO18. Our incomplete light curve is however consistent with this classification. Nevertheless the star is very faint and lies in the low main sequence. But membership analyses of RO18 and $\S$ \ref{membership} found the star to belong to the cluster.

V25. This star was classified by RO18 as a bright red variable or SR. However the star falls below the blue tail of the HB in our CMD (Fig. \ref{CMD}), consistent with the position found by RO18 in their CMD. The star's light curve presents a low amplitude sinusoidal variation with $P = 4.457$ d. The star lacks a proper motion measurement in the $Gaia$-DR2 which makes unclear its membership status. The star clearly cannot be a SR. Its proper classification is not clear.

V26. This clear sinusoidal variable resides to the red of the RGB in the CMD, as it has been consistently found by RO18 and us (Fig. \ref{CMD}), and both membership analyses found the star to be member. It has to be considered though, as it was warned in $\S$ \ref{membership}, that the cluster mean proper motion is not very different from that found in the field population, thus, V26 may be an example of a false positive detection. We tend to consider it not a cluster member. 

V31, V32, V33. Although these stars were given a variable star identification by RO18, their sinusuoidal variability was suspected by them. In our photometry we hardly detect any variation in V31 and V32 and do not confirm their variable status. On the other hand, V33 is in fact a sinusiodal variable with $P = 0.93353$ d. All three stars were found to be cluster members by RO18 and us ($\S$ \ref{membership}). The peculiar position of V33 in the bottom of the blue tail of the HB makes its classification uncertain.

V34. Due to the scarce number of observations in our work, we cannot confirm the variations reported by RO18. These authors were unable to classify the variability of the star, but considered it a cluster member. Unfortunately the Gaia source associated to this star has no proper motion information, but due to its peculiar position in the CMD we doubt its membership in the cluster.

N3. This is clearly a RRc star. Its position in the VPD and the membership determination approach ($\S$ \ref{membership}) make its belonging to the cluster dubious. However its position on the CMD nearly 2 mag below the HB  indicate the star is a field star behind the cluster. We performed the Fourier light curve decomposition and applied the calibrations described in $\S$ \ref{sec:V22} for V22, to estimate its iron abundance and distance. We found  
[Fe/H]$_{C09}=-1.11 \pm 0.19$ and d=10.3 kpc, confirming it as more distant object.

N1, N2, N4, N5 and N6. The variability and the status as no cluster members of these stars, were found by RO18. Although they are not in the FoV of our images, we included them in the membership analysis and conclude, like RO18, that they do not belong to the cluster.

% Don't change these lines
\bsp	% typesetting comment
\label{lastpage}

\end{document}